\documentclass[aip, reprint]{revtex4-1}
\usepackage{bm}
\usepackage{xcolor}
\usepackage{graphicx}
\usepackage{amsmath} 
\usepackage{url}
\usepackage{tikz}
\usepackage{braket}
\usepackage{comment}
\usepackage{xspace}
\usepackage[normalem]{ulem}
\usetikzlibrary{arrows,shapes,positioning,shadows,trees}
\usepackage{etoolbox}

\newcommand{\R}{\mathbf{R}}
\newcommand{\PyQMC}{\texttt{PyQMC}\xspace}
\newcommand{\PySCF}{\texttt{PySCF}\xspace}
\newcommand{\futures}{\texttt{futures}\xspace}
\newcommand{\highlight}[1]{\colorbox{red!20}{$\displaystyle#1$}}
\DeclareMathOperator*{\argmin}{arg\,min}
\definecolor{fig71}{HTML}{D0C7F0}
\definecolor{fig72}{HTML}{F0D0C7}
\definecolor{fig73}{HTML}{C7F0D0}
\usepackage{listings}
\usepackage{xcolor}

\definecolor{codegreen}{rgb}{0,0.6,0}
\definecolor{codegray}{rgb}{0.5,0.5,0.5}
\definecolor{codepurple}{rgb}{0.58,0,0.82}
\definecolor{backcolour}{rgb}{0.95,0.95,0.92}

\lstdefinestyle{mystyle}{
    backgroundcolor=\color{backcolour},   
    commentstyle=\color{codegreen},
    keywordstyle=\color{magenta},
    numberstyle=\tiny\color{codegray},
    stringstyle=\color{codepurple},
    basicstyle=\ttfamily\footnotesize,
    breakatwhitespace=false,         
    breaklines=true,                 
    captionpos=b,                    
    keepspaces=true,                 
    numbers=left,                    
    numbersep=5pt,                  
    showspaces=false,                
    showstringspaces=false,
    showtabs=false,                  
    tabsize=2
}

\makeatletter
\def\@email#1#2{%
 \endgroup
 \patchcmd{\titleblock@produce}
  {\frontmatter@RRAPformat}
  {\frontmatter@RRAPformat{\produce@RRAP{*#1\href{mailto:#2}{#2}}}\frontmatter@RRAPformat}
  {}{}
}%
\makeatother
\begin{document}

\title{\PyQMC: an all-Python real-space quantum Monte Carlo module in \PySCF}
\author{William A. Wheeler}
\affiliation{Department of Materials Science and Engineering, University of Illinois at Urbana-Champaign, Urbana, Illinois 61801, USA}
\author{Shivesh Pathak}
\affiliation{Center for Computing Research, Sandia National Laboratories, Albuquerque, New Mexico 87123, USA} 
\author{Kevin Kleiner}
\affiliation{Department of Physics, University of Illinois at Urbana-Champaign, Urbana, Illinois 61801, USA} 
\author{Shunyue Yuan}
\affiliation{Department of Applied Physics and Materials Science, California Institute of Technology, Pasadena, California 91125, USA} 
\author{Jo\~ao N. B. Rodrigues}
\affiliation{Centro de Ci\^encias Naturais e Humanas, Universidade Federal do ABC - UFABC, Santo Andr\'e, S\~ao Paulo 09210-580, Brasil}
\author{Cooper Lorsung}
\affiliation{Department of Mechanical Engineering, Carnegie Mellon University, Pittsburgh, PA 15213, USA}
\author{Kittithat Krongchon}
\affiliation{Department of Physics, University of Illinois at Urbana-Champaign, Urbana, Illinois 61801, USA} 
\author{Yueqing Chang}
\affiliation{Department of Physics and Astronomy, Rutgers University, Piscataway, New Jersey 08854, USA}
\author{Yiqing Zhou}
\affiliation{Laboratory of Atomic and Solid State Physics, Cornell University, Ithaca, New York 14853, USA}
\author{Brian Busemeyer}
\affiliation{Millenium Management, New York, New York 10022, USA}
\author{Kiel T. Williams}
\affiliation{Dynata, Plano, Texas 75024, USA}
\author{Alexander Mu\~noz}
\affiliation{Department of Physics, University of Illinois at Urbana-Champaign, Urbana, Illinois 61801, USA} 
\author{Chun Yu Chow}
\affiliation{Department of Physics, University of Illinois at Urbana-Champaign, Urbana, Illinois 61801, USA} 
\author{Lucas K. Wagner}
\affiliation{Department of Physics, University of Illinois at Urbana-Champaign, Urbana, Illinois 61801, USA}
\email[]{lkwagner@illinois.edu}

\date{\today}

\begin{abstract}
    We describe a new open-source Python-based package for high accuracy correlated electron calculations using quantum Monte Carlo (QMC) in real space: \PyQMC. 
    \PyQMC implements modern versions of QMC algorithms in an accessible format, enabling algorithmic development and easy implementation of complex workflows. 
	Tight integration with the \PySCF environment allows for simple comparison between QMC calculations and other many-body wave function techniques, as well as access to high accuracy trial wave functions. 
\end{abstract}
\maketitle

\section{Introduction}

\textit{Ab initio} calculations play an integral role in advancing our knowledge of molecules and materials.
They link materials properties to physical mechanisms in pristine systems, eliminating many difficult-to-control experimental factors.
Without the need for experimental inputs, \textit{ab initio} calculations and models also accelerate the search and design of new materials.\cite{lebegue_two-dimensional_2013, curtarolo_high-throughput_2013}
Strongly correlated materials, including unconventional superconductors,\cite{moree_ab_2022} 2D materials,\cite{choudhary_high-throughput_2017, wilson_excitons_2021} and defect systems,\cite{gali_ab_2019, dreyer_first-principles_2018} require computational approaches with careful treatment of electron correlation.\cite{adler_correlated_2018}

Calculations have an inherent trade-off between accuracy and computational cost: more accurate methods scale more steeply with number of electrons, and fully accurate calculations scale exponentially with system size.
Quantum Monte Carlo (QMC) offers a good balance between accuracy and scalability, capable of treating systems with thousands of electrons.\cite{foulkes_quantum_2001, martin_interacting_2016, wagner_discovering_2016, needs_variational_2020}
The past few years have seen several advances in QMC methods: new wave functions using machine learning techniques,\cite{pilati_self-learning_2019, pfau_abinitio_2020, acevedo_vandermonde_2020, hermann_deep-neural-network_2020, liInitioCalculationReal2022, wilson_simulations_2021} 
new algorithms for optimizing excited states,\cite{shea_size_2017, dash_excited_2019, otis_hybrid_2020, tran_improving_2020, feldt_excited-state_2020, dash_tailoring_2021, pathak_excited_2021}
complex observables such as energy density,\cite{krogel_quantum_2013, ryczko_machine_2022} and density matrices,\cite{wagner_types_2013} 
a new method to derive effective Hamiltonians from \emph{ab initio} QMC,\cite{changlani_density_2015, zheng_from_2018, chang_effective_2020}
and new time-stepping algorithms to reduce timestep error.\cite{zen_boosting_2016, anderson_nonlocal_2021}

Developing new tools and expanding the reach of QMC-level accuracy are necessary to address current problems in condensed matter physics, but comes with challenges.
Achieving highest performance can depend on subtle details of algorithm implementation,\cite{anderson_nonlocal_2021}
and adding new methods can require significant changes to algorithms.
A bottleneck in this development process is the testing and implementation of new ideas in code.
Several high-performance real-space QMC codes are under active development, including \texttt{QMCPACK},\cite{kim_qmcpack_2018} \texttt{CASINO},\cite{needs_variational_2020} \texttt{TurboRVB},\cite{nakano_turborvb_2020} and \texttt{CHAMP}.\cite{umrigar_cornell_nodate}
These real-space QMC software packages are written in low-level compiled languages such as C++ and/or Fortran\cite{umrigar_cornell_nodate, kim_qmcpack_2018, needs_variational_2020, nakano_turborvb_2020} to achieve high performance suitable for large-scale calculations; however, these packages are bulky (many lines of code) and challenging to modify.

To streamline development and teaching of new ideas in quantum Monte Carlo, we have written \PyQMC, an all-Python, flexible implementation of real-space QMC for molecules and materials.
\PyQMC is part of the \PySCF ecosystem, a collection of libraries that achieve performance close to that of compiled languages while being implemented in the much more flexible Python language. 
In this manuscript, we will describe the implementation of \PyQMC and note some of its advantages: integration with \PySCF, fast development, modularity and compatibility with user-modified code, flexibility of parallelization across diverse platforms (traditional desktop, cloud, high performance computing), and a unified codebase for running on graphics processing units or central processing units.

\section{QMC implementation}\label{sec:implementation}

There are many resources that offer thorough introductions to real-space QMC methods.\cite{foulkes_quantum_2001, hammond_monte_1994, nightingale_quantum_1998, prigogine_new_2009, kolorenc_applications_2011, austin_quantum_2012, toulouse_chapter_2016, wagner_discovering_2016, martin_interacting_2016}
Here, we will desribe our implementation of these methods in \PyQMC.

\subsection{Flexible wave functions}\label{sec:wfs}

\begin{figure}
	\includegraphics{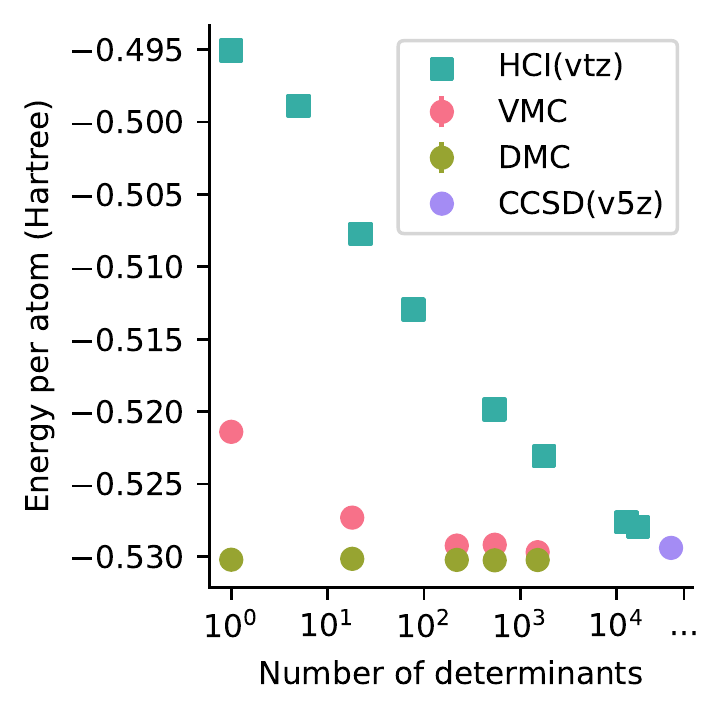}
    \caption{Ground state energy of 6-atom hydrogen chain at spacing of 3 Bohr. Because the nodal error is small, DMC performs well with only a single determinant. VMC with a two-body Jastrow achieves a similar result with an order of magnitude fewer determinants than the pure multi-determinant methods. This figure is reproduced from Ref.~\onlinecite{yuan_quantification_2022}.}
\label{fig:ndets}
\end{figure}

Wave functions are represented as Python objects in \PyQMC.
The standard implementation is the multi-Slater Jastrow (MSJ) wave function, having the form
\begin{equation}
\Psi(\R)  =  e^{J(\R, \bm{\alpha})} \sum_k c_k D_k(\R, \bm{\beta}),
\label{eqn:msj}
\end{equation} 
where $\R$ represents the positions of all the electrons, and $\bm{\alpha}$ (Jastrow), $\mathbf{c}$ (determinant), and $\bm{\beta}$ (orbital) are variational parameters.
Each determinant $D_k = \det \lbrace\phi_i^k(\mathbf{r}_j)\rbrace$ is constructed from a different set of single-particle orbitals $\lbrace\phi_i\rbrace$, where $\mathbf{r}_j$ is the position of electron $j$.
The two-body Jastrow,
\begin{equation}
J(\R; \bm{\alpha}) = \sum_{(i, j), I} u(r_{ij}, r_{iI}; \alpha),
\end{equation} 
 is a function of all the electron-electron ($r_{ij}$) and electron-nucleus ($r_{iI}$) distances, 
where $u$ is a function describing the cusp conditions and short-range correlation, defined in Ref~\cite{wagnerEnergeticsDipoleMoment2007}.
These wave functions are compatible with both open and twisted boundary calculations.

The MSJ trial function allows for a compact representation of the wave function by using fewer determinants to represent static correlation and the Jastrow factor to represent dynamical correlation.\cite{umrigar_optimized_1988}
Fig.~\ref{fig:ndets} compares the number of determinants needed with and without a Jastrow factor for a chain of six hydrogen atoms with lattice spacing 3.0 Bohrs, in the strongly correlated regime. 
The variational Monte Carlo (VMC) calculation uses a two-body Jastrow with electron-electron and electron-ion pair correlation.
Fixed-node diffusion Monte Carlo (DMC) can be interpreted as using the best possible Jastrow factor, shifting the wave function distribution without changing the nodal surface.
Coupled cluster (CCSD) in the V5Z basis is near the complete basis set limit, and is consistent with the DMC energy.
Heat-bath configuration interaction (HCI) approaches the CCSD value as the determinant basis size increases.
The pure determinant methods require two orders of magnitude more determinants than the VMC with a two-body Jastrow to converge.

In addition, we have implemented \texttt{J3}, the three-body Jastrow proposed by Sorella et. al.,\cite{sorella_weak_2007} for open boundary calculations.
Any number of wave functions can be combined through the \texttt{MultiplyWF} and \texttt{AddWF} objects, enabling mixing and matching of wave function forms.
For efficiency, the Slater object includes a linear combination of determinants without the need for combining multiple wave function objects.
As a subject of active research, we expect additional wave function forms to be added over time.

New wave functions are easily implemented in the \PyQMC framework.
Any object that conforms to the wave function interface can be used in all \PyQMC methods.
For example, other groups have implemented neural network trial functions\cite{liInitioCalculationReal2022} and used the algorithm outlined in section \ref{sec:vmc} to optimize the wave function parameters.
\PyQMC's testing framework makes it possible to quickly check for bugs and ensure compatibility of new objects for seamless integration.

\subsection{Expectation values of arbitrary operators}

An arbitrary operator $\hat{O}$ is evaluated on wave functions $\Phi$ and $\Psi$ as follows
\begin{align}\label{eq:operator_evaluation}
	\frac{\braket{\Phi|\hat{O}|\Psi}}{\braket{\Phi|\Psi}} &= \frac{\int d\R d\R' \Phi^*(\R) \Psi(\R')  O(\R,\R')}{\int{d\R \Phi^*(\R)\Psi(\R)}} \\
    &= \frac{\int d\R  \Phi^*(\R) \Psi(\R) \highlight{\int d\R' \frac{\Psi(\R')}{\Psi(\R)} O(\R,\R')}}{\int{d\R \Phi^*(\R)\Psi(\R)}} \\
    &= \frac{\int d\R  \Phi^*(\R) \Psi(\R) \highlight{O_L(\R,\Psi)}}{\int{d\R \Phi^*(\R)\Psi(\R)}}
\end{align}
where the highlighted term is the local evaluation of the operator $\hat{O}$, 
\begin{equation}
    O_L(\R, \Psi) = \int{d\R' \frac{\Psi(\R') O(\R,\R')}{\Psi(\R)}  } .
\label{eq:nonlocal_operator}
\end{equation}
In \PyQMC, the integral over $\R$ is handled by the VMC algorithm, where $\Psi=\Phi=\Psi_T$, and in the case of DMC, $\Psi=\Psi_T$ is the trial function while $\Phi = \Phi_{FN}$ is the fixed-node wave function.
We define an accumulator as an object that evaluates $\mathcal{O}_L(\R, \Psi)$.

In this section, we summarize the accumulator objects implemented in \PyQMC.

\subsubsection{Gradient operators}

For semilocal operators such as gradients, the expression in Eq.~\ref{eq:nonlocal_operator} simplifies to
\begin{equation}
    \mathcal{O}_L(\R) = \frac{[\hat{O}\Psi](\R) }{\Psi(\R)}.
\end{equation}
In \PyQMC, all wave function objects can compute $\frac{\nabla_r \Psi}{\Psi}$, $\frac{\nabla_r^2 \Psi}{\Psi}$, and $\frac{\nabla_p \Psi}{\Psi}$, where $\nabla_r$ refers to the gradient with respect to a single electronic coordinate, and $\nabla_p$ refers to the gradient with respect to all variational parameters in the wave function. 

\subsubsection{Effective core potentials}


\PyQMC is compatible with semilocal effective core potentials (ECPs) (nonlocal in the angular part, but local in the radial part). 
ECP evaluation is implemented as in \texttt{QWalk}\cite{wagner_qwalk_2009} using the form described by Mitas et al.\cite{mitas_nonlocal_1993}
\PyQMC automatically reads the ECPs from the \PySCF Mole or Cell object.

The nonlocal operator takes the form of Eq.~\ref{eq:nonlocal_operator}. 
The ECP operator $H^{\rm ECP}(\R, \R') = \sum_{e,a} H_{ea}^{\rm ECP}(\R, \R')$ is a sum of independent terms between electron $e$ and atom $a$,
\begin{equation}
H_{ea}^{\rm ECP}(\R, \R') = \delta(r_{ea} - r_{ea}') \sum_l \frac{2l + 1}{4\pi} v_l(r_{ea}) P_l(\cos \theta'),
\end{equation}
where $r_{ea}$ is the distance between positions of electron $e$ and atom $a$, $v_l$ is a radial pseudopotential for angular momentum channel $l$, $P_l$ is a Legendre polynomial, and $\theta'$ is the angle between $\mathbf{r}_{ea}$ and $\mathbf{r}_{ea}'$.
The angular integral for each $(e, a)$ pair is evaluated using a randomly oriented quadrature rule
\begin{align*}
\int d\R' &H_{ea}^{\rm ECP}(\R, \R') \frac{\Psi(\R')}{\Psi(\R)} 
\\&= 
\frac{4\pi}{N_\Omega} \sum_{\Omega} w_\Omega H_{ea}^{\rm ECP}(\R, \R')\frac{\Psi(\R_{ea\Omega}')}{\Psi(\R)} ,
\end{align*}
where the auxiliary configurations $\R_{ea\Omega}'$ are generated from $\R$ by moving electron $e$ about ion $a$ by angles $\Omega = (\theta, \phi)$ of the quadrature grid and corresponding weights $w_\Omega$.
\PyQMC has implemented all the quadrature rules of octahedral and icosahedral symmetries listed by Mitas et. al.\cite{mitas_nonlocal_1993}

\subsubsection{Reduced density matrices}\label{sec:rdm}
\begin{figure}
    \includegraphics{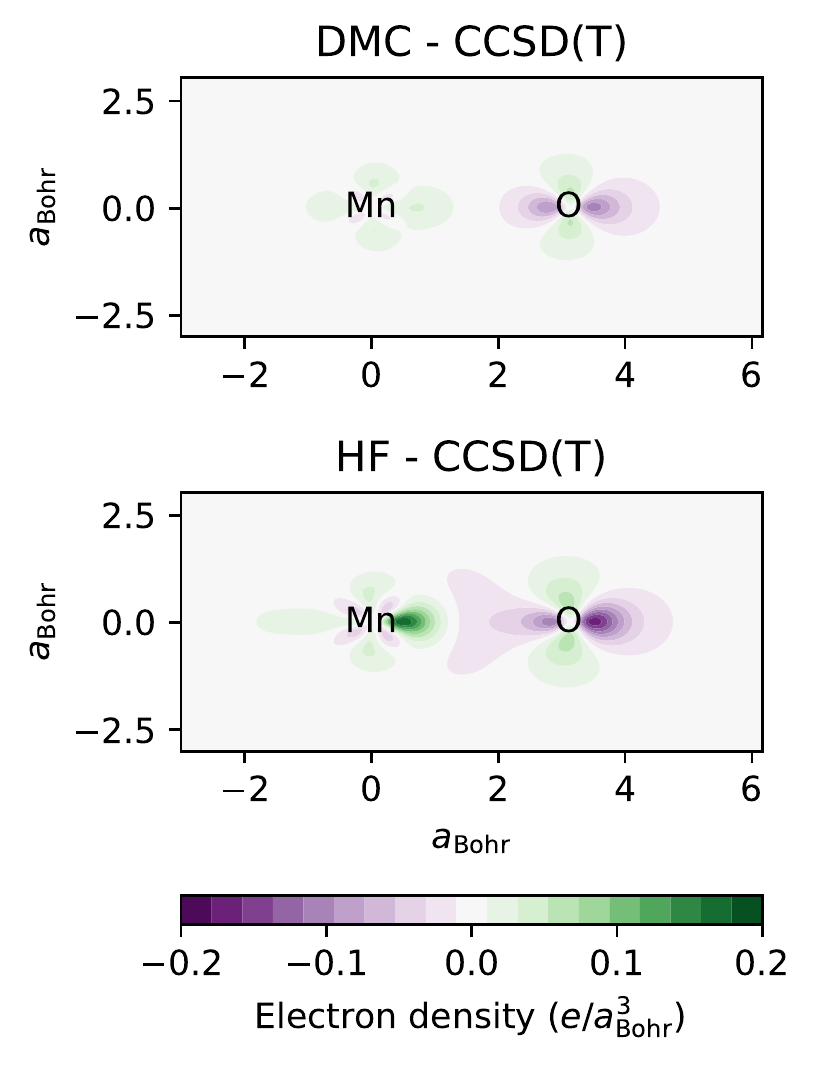}
    
\begin{lstlisting}[language=Python]
import numpy as np
import pyqmc.api as pyq
import pyscf.tools

mol, mf = pyq.recover_pyscf("MnO.chk")
rdm1_dmc = pyq.read_mc_output(
    "MnO_DMC.hdf5", 
    warmup=100, 
    reblock=20,
)
rdm1_ccsd_t = np.load("MnO_CCSD(T).npy")
ao_rdm1_dmc = np.einsum(
    "pi,ij,qj->pq", 
    mf.mo_coeff, 
    rdm1_dmc - rdm1_ccsd_t, 
    mf.mo_coeff.conj()
)
dens_dmc = pyscf.tools.cubegen.density(
    mol, "MnO_dmc.cube", ao_rdm1_dmc
)
\end{lstlisting}
\caption{The integration between \PySCF and \PyQMC makes it straightforward to compare properties of wave functions between different high level methods. 
(a) Electron density of the MnO molecule at bond length 1.6477~{\AA} for Hartree-Fock and diffusion Monte Carlo referenced to CCSD(T) calculations.
(b) The code used to compute the DMC densities. For the entire code used to generate the plots, see the supplementary information.
}
\label{fig:density_diff}
\end{figure}

All one-particle observables can be calculated from the one-particle reduced density matrix (1-RDM), making it a useful quantity to characterize many-body wave functions alongside the energy.
In \PyQMC, the 1-RDM is represented in an basis of single-particle orbitals $\phi_i(\mathbf{r})$ as
\begin{equation}
\rho_{ij} = \langle \Psi | c_i^\dagger c_j | \Psi \rangle,
\end{equation}
where $c_i^\dagger$ and $c_i$ are creation and annihilation operators for orbital $\phi_i$, respectively.

Since the reduced density matrices are completely nonlocal, we perform an auxilliary random walk, sampling a conditional probability $P'(\R'|\R)$ and evaluating 
\begin{equation}
    \mathcal{O}(\R) = \left\langle \frac{\Psi(\R') O(\R,\R')}{\Psi(\R)P'(\R'|\R)} \right\rangle_{\R' \sim P'(\R'|\R)}.
\end{equation}

The 1-RDM is evaluated in QMC by averaging the quantity\cite{wagner_types_2013}
\begin{equation}
\rho_{ij} = \frac{1}{\sqrt{N_iN_j}}\left\langle \sum_{a=1}^N 
\frac{\Psi^*(\R'_a)}{\Psi^*(\R)} 
\frac{\phi_i(\mathbf{r}'_a) \phi_j^*(\mathbf{r}_a)}{\rho_{\rm aux}(\mathbf{r}_a')} 
\right\rangle_{\begin{subarray}{l}\R\sim|\Psi|^2;\\ \mathbf{r}_a'\sim\rho_{\rm aux}\end{subarray}},
\end{equation} 
where $\R'_a$ is generated from $\R$ by moving electron $a$, $\mathbf{r}_a \rightarrow \mathbf{r}_a'$,
and $\rho_{\rm aux}(\mathbf{r}) =  \sum_i |\phi_i(\mathbf{r})|^2$ is proportional to the one-particle distribution used to sample the auxiliary coordinate $\mathbf{r}_a'$.
We use McMillan's method of using the same auxiliary coordinates $\mathbf{r}_a'$ for every electron $a$ in the sum.\cite{mcmillan_ground_1965}
The normalization factors
\begin{equation}
N_i = \left\langle \frac{|\phi_i(\mathbf{r})|^2}{\rho_{\rm aux}(\mathbf{r})} \right\rangle_{\mathbf{r}\sim\rho_{\rm aux}}
\end{equation}
 are accumulated during the Monte Carlo run, and are applied as a post-processing step using the function \texttt{normalize\_obdm}.

The two-particle reduced density matrix (2-RDM) 
\begin{equation}
\rho_{ijkl} = \langle \Psi | c_i^\dagger c_k^\dagger c_l c_j | \Psi \rangle
\end{equation} 
can be used to calculate all two-body observables, and is analogous to the 1-RDM.
Note that in some modules in \PySCF and other quantum chemistry codes, $\langle \Psi | c_i^\dagger c_j c_k^\dagger c_l | \Psi \rangle$ is evaluated instead. 
It is relatively easy to translate between these two representations as
\begin{equation}
\langle \Psi | c_i^\dagger c_k^\dagger c_l c_j | \Psi \rangle
=
\langle \Psi | c_i^\dagger c_j c_k^\dagger c_l | \Psi \rangle
- \delta_{jk} \langle \Psi | c_i^\dagger c_l | \Psi \rangle,
\end{equation} 
which is done by the \PySCF function \texttt{reorder\_rdm}.
The 2-RDM is evaluated in QMC as
\begin{equation}
\rho_{ijkl} = \left\langle \sum_{a<b} \frac{\Psi(\R'_{ab})}{\Psi(\R)} 
\frac{\phi^*_{j}(\mathbf{r}_a') \phi^*_{l}(\mathbf{r}_b') \phi_{i}(\mathbf{r}_a) \phi_{k}(\mathbf{r}_b) }{\rho_{\rm aux}(\mathbf{r}_a')\rho_{\rm aux}(\mathbf{r}_b')} 
\right\rangle_{\begin{subarray}{l}\R\sim|\Psi|^2;\\ \mathbf{r}_a'\sim\rho_{\rm aux}\end{subarray}},
\end{equation} 
\PyQMC's implementation can evaluate the RDMs in an arbitrary basis.

\PySCF routines can be applied directly to the 1-RDM computed in DMC to compute and plot density or other one-body quantities (Fig.~\ref{fig:density_diff}).
Using \PySCF's built-in \texttt{cubegen.density} function removes the need to write a new script for plotting.

Computing RDMs in the same basis allows for seamless comparison between methods, i.e., by simply subtracting the matrices.
Different methods are commonly compared by their energies, a single number.
One- and two-particle density matrices capture more of the state and offer better comparison of properties; methods that result in the same energy may still produce states with different densities.
QMC computations of RDMs in a basis have an additional advantage that the statistical noise is much smaller compared to computing on a grid, resulting in smoother density plots.
The difference in densities between Hartree-Fock, DMC, and CCSD(T) is shown in Fig.~\ref{fig:density_diff}.

\subsection{Bulk systems}

Infinite solids are approximated by finite simulation cells with twisted boundary conditions (TBCs)
\begin{equation}
    \Psi(\mathbf{r}_1, \ldots, \mathbf{r}_i + \mathbf{L}, \ldots, \mathbf{r}_N)
    =
    e^{i\mathbf{k}\cdot\mathbf{L}} \Psi(\mathbf{r}_1, \ldots, \mathbf{r}_i, \ldots, \mathbf{r}_N),
\end{equation} 
where $\mathbf{k}$ is the twist; thus the basis functions are eigenstates of a translation operator. 
The one-particle part of the Hamiltonian commutes with the translation of a single electron, and thus can be diagonalized using basis functions of definite twist.
The total energy per cell is obtained by averaging over all twists in the Brillouin zone.\cite{lin_twist-averaged_2001}
However, the Coulomb operator does not commute with the translation operator of a single electron, and thus causes the energy eigenstates to in general be superpositions of twists.

In \PyQMC, practical calculations are performed using a \textit{supercell} approximation, in which a simulation cell larger than the primitive cell is chosen, and the Coulomb operator is truncated to remove matrix elements between different twists. 
This truncation can be partially corrected using the structure factor,\cite{chiesaFiniteSizeErrorManyBody2006} with an error proportional to $\frac{1}{N}$, where $N$ is the number of electrons in three dimensions. 
Note that this correction should be performed after twist averaging above. 

\PyQMC contains several features to facilitate extrapolation to infinite system size. 
First, a \PySCF mean-field calculation is performed on the primitive cell. 
The $k$-points used in the mean-field calculation determine which twists are available for a given supercell $\mathbf{S}$.
The available twists are obtained in \PyQMC using the function \texttt{available\_twist(cell, mf, S)}, where \texttt{cell} and \texttt{mf} are \PySCF cell and mean-field objects. 
The code then automatically generates the appropriate supercell objects from the primitive cell mean-field object.
By averaging over twist, one can remove the kinetic energy finite size correction.\cite{chiesaFiniteSizeErrorManyBody2006}
The structure factor is available as an accumulator. 
The small-k limit of the structure factor gives the approximate Coulomb finite size correction.\cite{chiesaFiniteSizeErrorManyBody2006}

\subsection{Methods}

\subsubsection{Variational Monte Carlo (VMC)}\label{sec:vmc}

The trial functions in section~\ref{sec:wfs} can contain hundreds or thousands of parameters. 
To approximate the ground state, the parameters of the trial function are variationally optimized by minimizing the VMC energy
\begin{equation}
E[\Psi] = \langle \Psi | \hat{H} | \Psi \rangle = \left\langle \frac{\hat{H}\Psi (\R)}{\Psi(\R)} \right\rangle_{R\sim|\Psi|^2}.
\end{equation}

\begin{figure}
\begin{enumerate}
\item Generate walkers $\R$
\item Compute regularization factor\cite{pathak_light_2020} $f(\R)$
\begin{enumerate}
\item $d(\R) \leftarrow \frac{1}{r_{\rm cutoff}}\frac{\Psi(\R)}{\sqrt{\sum_e |\nabla_e \Psi(\R)|^2}}$
\item $f(\R) \leftarrow  9d(\R)^2 - 15d(\R)^4 + 7d(\R)^6$
\end{enumerate}
\item Stochastic reconfiguration\cite{casula_geminal_2003}
\begin{enumerate}
\item $G_{\Psi}^i(\R) \leftarrow \frac{\partial_{p_i} \Psi}{\Psi}\big\vert_\R f(\R)$
\item $E_L(\R) \leftarrow \frac{\hat{H}\Psi}{\Psi} \big\vert_{\R}$
\item $G_E^i \leftarrow \braket{E_L(\R) G_\Psi^i(\R)}_\R - \braket{E_L(\R)}_\R \braket{G_\Psi^i(\R)}_\R$
\item $S_{ij} \leftarrow \left\langle  \frac{\partial_{p_i}\Psi}{\Psi}\big\vert_\R G_\Psi^j(\R) \right\rangle_\R  - \left\langle  G_\Psi^i(\R) \right\rangle_\R \left\langle G_\Psi^j(\R) \right\rangle_\R$
\item $ u_i \leftarrow \sum_j (\mathbf{S})^{-1}_{ij} G_E^j \text{\, (regularized gradient)}$
\end{enumerate}
\item Line minimization using correlated sampling
\begin{enumerate}
\item Select walkers $\R$
\item $\mathbf{p} \leftarrow \text{ parameters of } \Psi$
\item \textbf{for} $x$ in [-1, 0, 1, 2, 3], \textbf{do}
\begin{enumerate}
\item $ \Psi_x \leftarrow \text{ replace } \mathbf{p} \text{ with } \mathbf{p} + x\mathbf{u}$
\item $ E(x) = \left\langle \frac{\hat{H}\Psi_x}{\Psi_x} \left|\frac{\Psi_x}{\Psi_0}\right|^2 \right\rangle_{\R}$
\end{enumerate}
\item $E_{\rm fit} \leftarrow \text{fit } E(x) \text{ to cubic function}$
\item $ x_{\rm min} \leftarrow \argmin_x\, E_{\rm fit}(x)$
\item $\Psi \leftarrow \Psi_{x_{\rm min}}$
\end{enumerate}
\end{enumerate}
\caption{Pseudo-code for the wave function optimization routine in \PyQMC. The three main parts of each step of the optimization algorithm: variance regularization factor, stochastic reconfiguration, and line minimization. }
\label{fig:optimization_algorithm}
\end{figure}

The gradient of $E[\Psi]$ is used to determine the updates to the parameters $\mathbf{p}$ during optimization.
The gradient estimator $\frac{\hat H \Psi}{\Psi} \frac{\partial_p \Psi}{\Psi}$ has infinite variance near the nodes of $\Psi$, which is removed by including the regularization factor of Ref~\cite{pathak_light_2020}.
Next, the parameter update direction is determined from $\frac{\partial E}{\partial p}$ using the stochastic reconfiguration technique of Casula and Sorella.\cite{casula_geminal_2003}
Finally, the magnitude is determined by the minimum energy along the update direction. 
The parameters corresponding to the minimum are determined by a polynomial fit of correlated samples of the energy along the line.
The parameters are updated, and the process is repeated to convergence.

For multi-Slater-Jastrow functions (Eq.~\ref{eqn:msj}), \PyQMC supports optimization of $\alpha$ (Jastrow), $c$ (determinant), and $\beta$ (orbital) parameters.

\subsubsection{VMC for excited states}

\begin{figure}
    \includegraphics[width=0.5\textwidth]{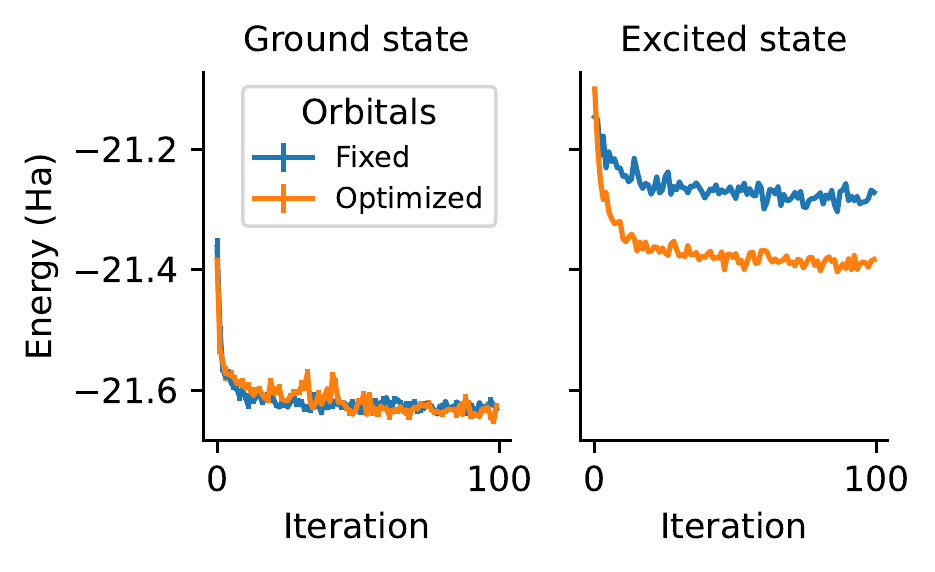}
    \caption{Optimization of a minimal multi-Slater-Jastrow wave function for the ground and first excited states of a CO molecule, with and without optimizing the orbitals.
Better variational estimates are achieved for the excited state by optimizing the orbitals.}
\label{fig:orbital_optimization}
\end{figure}
    
A standard approach to computing excited states is to hold orbital coefficients fixed from an excited mean-field determinant.\cite{williamson_diffusion_1998}
To optimize excited-state orbitals, additional measures are required to keep them from reverting to the orbitals of lower-energy states.
Methods such as the state-averaged CASSCF method\cite{docken_lih_1972, werner_second_1985} or other state averaged methods\cite{schautz_optimized_2004, filippi_absorption_2009, dash_excited_2019, cuzzocrea_variational_2020, dash_tailoring_2021} allow orbital shapes to vary, but require the orbitals to be the same for all energy eigenstates. 
This requirement makes it easier to enforce orthogonality of eigenstates but severely limits the expressiveness of the wave functions.

In \PyQMC's implementation, excited states are kept orthogonal to lower-energy states through an overlap penalty introduced in Ref~\cite{pathak_excited_2021}, allowing orbital coefficients to be optimized for each state independently.
The objective function for the optimization is given by 
\begin{align}
    O[\Psi] &=\langle \Psi | \hat{H} | \Psi \rangle + \sum_{i=1}^{n-1} \lambda_i |\langle \Psi_i | \Psi \rangle|^2 \\
            &=\left\langle \frac{\hat{H}\Psi (\R)}{\Psi(\R)} \frac{|\Psi(\R)|^2}{\rho} \right\rangle 
             + \sum_{i=1}^{n-1} \lambda_i \left| \frac{N_{in}}{\sqrt{N_{ii}N_{nn}}} \right|^2,
\end{align}
where 
\begin{equation}
N_{ij} = {\left\langle \frac{\Psi_i^*(\R)\Psi_j(\R)}{\rho(\R)}\right\rangle},
\end{equation}
is the wave function overlap matrix and $\R$ is sampled from the distribution $\rho(\R)$.
Typically $\rho(\R) \propto \sum_i |\Psi_i(\R)|^2$. 

To demonstrate the importance of orbital optimization for excited states, we show optimizations of the ground and first excited states of the CO molecule (Fig.~\ref{fig:orbital_optimization}) using a 400-determinant multi-Slater-Jastrow ansatz, with and without optimizing orbitals.
The energy is shown at each iteration over the course of both optimizations.
Fixed-orbital wave functions yield an excitation energy of 9.43(5) eV, compared with 6.68(5) eV from optimized-orbital wave functions of the same form.
Compared with the experimentally determined vertical excitation energy 4.76 eV,\cite{tobias_potential_1960} optimizing orbitals results in a 60\% improvement at the VMC level.

\subsubsection{Diffusion Monte Carlo}

Diffusion Monte Carlo (DMC) is implemented in \PyQMC using importance sampling and the fixed-node approximation.
Samples are drawn from the mixed distribution 
$f(\R) = |\Psi^*(\R)\Phi_0(\R)|$,
 where $\Phi_0$ is the ground state,  by stochastically applying a projection operator $\hat{\mathcal{P}}_\tau$ to a trial function $\Psi$,
\begin{equation}
\Psi^*(\R)\Phi_0(\R) = \lim_{N\rightarrow\infty} \braket{\Psi|\R}\braket{\R | \hat{\mathcal{P}}_\tau^N | \Psi}.
\end{equation} 
The time step $\tau$ is a parameter that must be extrapolated to $\tau\rightarrow 0$.
Positions and weights $(\R_i, w_i)$ are generated by the projection $\hat{\mathcal{P}}_\tau$ at each Monte Carlo step.
The fixed-node approximation is used for real wave functions, rejecting moves $\R\rightarrow\R'$ that change the sign of the trial function $\Psi$.
For complex wave functions, the fixed-phase approximation is used.\cite{ortiz_new_1993}
Because the gradient of the phase enters into the potential, no rejection based on sign is required.

Sampling the mixed distribution results in mixed-estimator averages
\begin{equation}
\langle \hat{O} \rangle = \frac
{\langle \Psi | \hat{O} | \Phi_0\rangle}
{\langle \Psi | \Phi_0\rangle}
\label{eq:mixed_estimator}
\end{equation}
which are computed similarly to Eq.~\ref{eq:operator_evaluation} as averages over walkers with additional weights $w_i$,
\begin{equation}
\left\langle w_i \frac{\int d\R'\, O(\R_i, \R') \Psi^*(\R')}{\Psi^*(\R_i)} \right\rangle_{\R_i\sim f(\R)}.
\end{equation} 

Branching is performed every few steps to keep weights balanced, replicating some walkers and removing others depending on their weights.
In \PyQMC, the branching is implemented by the stochastic comb method.\cite{assaraf_diffusion_2000, buonaura_numerical_1998, davis_critical-size_1961}
where walkers are resampled with probability proportional to their weights, the total weight $\sum_j w_j$ is saved, and the new weights are subsequently set equal to one.
The expected contribution from each walker is correct on average, and the resulting population bias is small.
This approach has the advantage of keeping the number of walkers fixed, which simplifies efficient parallelization on a fixed number of processors.

\PyQMC employs two strategies proposed by Anderson and Umrigar\cite{anderson_nonlocal_2021} to reduce time-step errors:
modified weight updates and modified T-moves\cite{casula_beyond_2006} for nonlocal ECPs.

\section{Diverse workflow support} \label{sec:workflow}

\subsection{Integration with \PySCF}

In many QMC codes, converters from other packages make up a large portion of the programming effort.
\PyQMC uses \PySCF objects directly to initialize calculations, eliminating the need for converters.
Mole and Cell objects define the Hamiltonian, including geometry, number of electrons, basis set, and pseudopotentials.
The use of \PySCF's \texttt{eval\_gto()} function to evaluate orbitals guarantees compatibility with any basis set supported by \PySCF.
QMC trial wave function determinants are generated from SCF objects, and there is some compatibility with multireference methods such as CAS, CASSCF, and full CI without requiring conversion steps. 

Tight coupling to \PySCF enables easy use of analysis routines.
A common example is the calculation and plotting of density differences discussed in section \ref{sec:rdm} and shown in Fig.~\ref{fig:density_diff}.

\PyQMC allows for file-free computation -- executing a full calculation from atomic structure to QMC result without saving any intermediate results (Fig.~\ref{fig:file-free}).
Having all objects and data in the workspace streamlines prototyping of new algorithms and workflows.

\begin{figure}
\begin{lstlisting}[language=Python]
from pyscf import gto, scf
import pyqmc.api as pyq

mol = gto.M(
    atom=f"Mn 0. 0. 0.; O 0. 0. 1.6477",
    basis="ccecp-ccpvtz",
    ecp="ccecp",
    spin=5,
)
mf = scf.UHF(mol)
mf.run()

# QMC
configs = pyq.initial_guess(mol, nconfig)
wf, to_opt = pyq.generate_wf(mol, mf)
pgrad_acc = pyq.gradient_generator(mol, wf, to_opt)
wf, optimization_data = pyq.line_minimization(wf, configs, pgrad_acc)
configs, dmc_data = pyq.rundmc(wf, configs)
\end{lstlisting}
\caption{Single script execution of a QMC calculation from atomic positions to QMC result. 
\PySCF objects are used directly in \PyQMC functions.
No writing intermediate results to disk is required.
}
\label{fig:file-free}
\end{figure}

\subsection{Monkey patching}\label{sec:monkey_patching}

\begin{figure}
\begin{lstlisting}[language=Python]
import numpy as np
import pyqmc.api as pyq


class DipoleAccumulator:
    def __init__(self):
        pass

    def __call__(self, configs, wf):
        return {"electric_dipole": configs.configs.sum(axis=1)}

    def avg(self, configs, wf):
        avg = {}
        data = self(configs, wf)
        for k, it in data.items():
            avg[k] = np.mean(it, axis=0)
        return avg

    def shapes(self):
        return {"electric_dipole": (3,)}

    def keys(self):
        return self.shapes().keys()


accumulators = {"extra_accumulators": dict(dipole=DipoleAccumulator())}
pyq.VMC(
    "MnO_scf.hdf5", 
    "MnO_vmc_dipole.hdf5", 
    load_parameters="MnO_opt.hdf5", 
    accumulators=accumulators,
)
\end{lstlisting}
\caption{User code can be injected into \PyQMC's QMC routines. In this example, we defined a class on the fly to compute the molecular dipole moment within the script that performs the calculation.}
\label{fig:dipole_accumulator}
\end{figure}

\begin{figure}
\begin{tikzpicture}[
    >=latex, 
    group/.style={fill=fig71, rounded corners=6pt, minimum height=5cm, minimum width=4cm, drop shadow},
    basic/.style  = {draw, text width=2cm, drop shadow, font=\sffamily, rectangle},
    box/.style={basic, rounded corners=3mm, draw, fill=fig72},
    level 2/.style = {rounded corners=4pt, align=left, text width=10em, xshift=1mm, font=\sffamily},
    level 3/.style = {basic, thin, align=left, fill=fig73, text width=9em},
    level 3wf/.style = {basic, thin, align=left, fill=fig73, text width=6em},
    level 1/.style={sibling distance=40mm},
    edge from parent/.style={->,draw}5
]

\node(wfs) [group, minimum width=3.2cm]{};
\node(wfslabel)[level 2, below=2mm of wfs.north west, anchor=north west]{\large Wave functions};
\node(slater)[level 3wf, below=2mm of wfslabel, xshift=-0.5em]{Multi-Slater};
\node(jastrow)[level 3wf, below=2mm of slater]{Jastrow};
\node(j3)[level 3wf, below=2mm of jastrow]{J3};
\node(user)[level 3wf, below=2mm of j3]{User-defined};

\node(accs) [group, right=1cm of wfs]{};
\node(accslabel)[level 2, below=2mm of accs.north west, anchor=north west]{\large Accumulators};
\node(energy)[level 3, below=2mm of accslabel, xshift=1em]{Energy};
\node(pgrad)[level 3, below=2mm of energy]{Parameter gradients};
\node(Sq)[level 3, below=2mm of pgrad]{Structure factor};
\node(1rdm)[level 3, below=2mm of Sq]{1-RDM};
\node(2rdm)[level 3, below=2mm of 1rdm]{2-RDM};
\node(user)[level 3, below=2mm of 2rdm]{User-defined};

\node(vmcdmc) [box, below right = 1cm and 5mm of wfs.south east, anchor=north, minimum width=4cm, minimum height=1cm]{\centering \large VMC/DMC};

\node(output) [box, below=1cm of vmcdmc, minimum width=4cm, minimum height=1cm]{\parbox{2cm}{\centering \large Output observables}};


\node(joinarrows) [coordinate, above=6mm of vmcdmc.north]{};
\draw[->, very thick] (accs.south) -- (accs.south |- joinarrows) -- (joinarrows) -- (vmcdmc.north);
\draw[->, very thick] (wfs.south) -- (wfs.south |- joinarrows) -- (joinarrows) -- (vmcdmc.north);
\draw[->, very thick] (vmcdmc) -- (output);
\end{tikzpicture}
\caption{User-defined code can be mixed and matched in multiple ways. (a) Externally defined accumulators are input directly into built-in VMC and DMC routines as in Fig~\ref{fig:dipole_accumulator}. (b) User-defined custom VMC algorithms outside of the \PyQMC package are run using the built-in \PyQMC accumulators and wave functions.}
\label{fig:flowchart_vmc}
\end{figure}
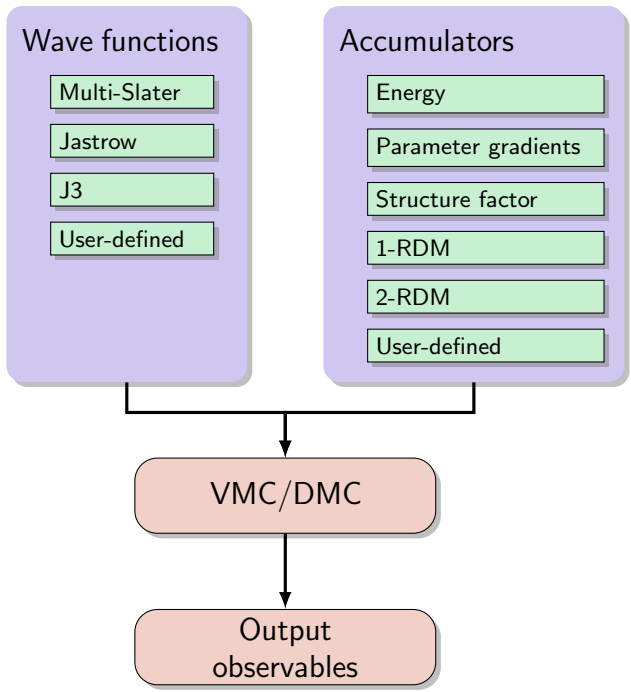

\PyQMC allows users to add modifications to a calculation locally without changing the package directory, a practice known as ``monkey patching.''
Although modifying the package directory is certainly possible, it poses a barrier to users in our experience.
With \PyQMC's all-Python, modular structure, built-in routines are compatible with objects defined outside the package directory, such as customized wave function and accumulator objects for VMC and DMC;
built-in objects can be used in externally-defined customized methods as well.
Fig.~\ref{fig:dipole_accumulator} shows code outside of the package defining an accumulator object that is used directly in \PyQMC's VMC and DMC routines, in this case to compute the dipole moment of a molecule.
Using custom accumulator objects is depicted in the flowchart in Fig.~\ref{fig:flowchart_vmc}. 

As an example of the benefits of this platform, we contrast implementation of a new VMC algorithm between Python and C++ (e.g. for sampling the sum of two wave functions in excited state optimizations). 
In C++, the new algorithm would require adding a file into the package, adding the file into the make system, and recompiling the distribution. 
In Python, a customized VMC is written, tested, and run at scale without the user modifying the distributed package at all, as depicted in the flowchart in Fig.~\ref{fig:flowchart_vmc}.
It is completely portable; the new algorithm file(s) can be shared and it will work for another user or machine. 
Developing new QMC methods and algorithms is often iterative, and by requiring fewer steps, this Python implementation greatly reduces friction for users and developers to explore new ideas.

\section{Acceleration strategies}

\PyQMC supports two acceleration strategies: parallel execution, and the use of graphical processing units (GPUs). 
The strategies work simultaneously: quantum Monte Carlo calculations can use multiple GPUs across multiple computational resources.
It is possible to parallelize on heterogenous resources, in which some calculations are performed on CPUs and some on GPUs.

\subsection{Parallelization}

\PyQMC makes use of Python's standard library \futures objects for parallelization.
For compatibility with \PyQMC, a futures object need only implement the \texttt{submit} function, which distributes work onto a remote process or server. 
The wave function data and a subset of walkers are sent to each worker process, and the results are collected as the processes finish.

By using \futures objects, \PyQMC can transparently take advantage of many parallelization strategies.
The Python standard library \texttt{concurrent.futures} provides on-node process-based parallelization. 
Other packages can be installed and used with the code transparently; for example \texttt{mpi4py}\cite{dalcin_mpi_2005, dalcin_mpi_2008, dalcin_parallel_2011, dalcin_mpi4py_2021} provides futures over the high performance computing standard Message Passing Interface.\cite{mpi40}
Similarly, \texttt{Dask}\cite{dask} provides a futures-based interface using pilot processes that are very flexible, allowing for remote execution on cloud-based resources. 

\begin{figure}
    \includegraphics{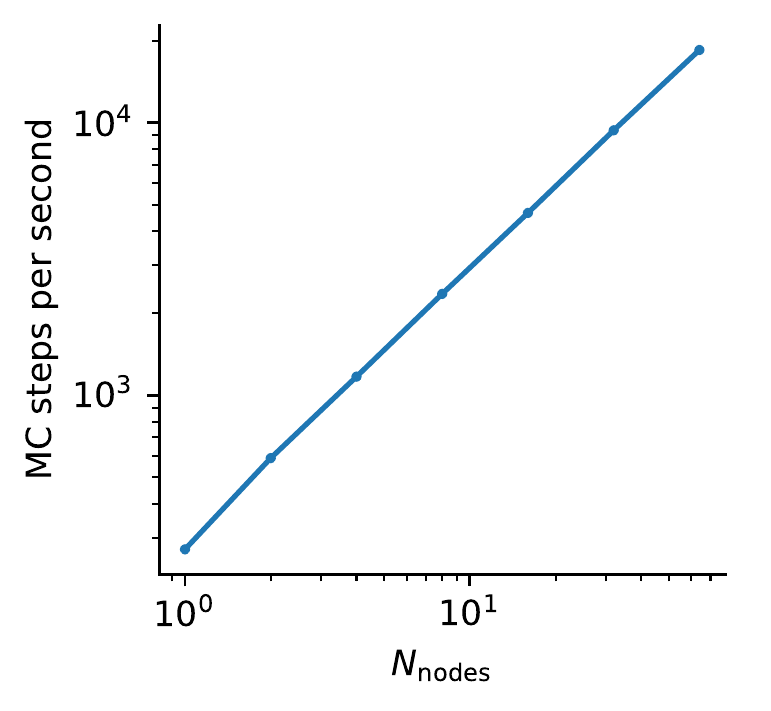}
    \caption{Parallel scaling of VMC on a coronene molecule, with the number of walkers scaled proportionally to the number of cores. Calculations were run on Summit and parallelized with \texttt{Dask}.\cite{dask} }
\label{fig:parallel_speedup}
\end{figure}
    
Quantum Monte Carlo methods are often termed ``embarrassingly parallel,'' meaning that the computational time decreases almost linearly as the number of processors increases.
Fig.~\ref{fig:parallel_speedup} shows the number of Monte Carlo steps executed per second as a function of the number of nodes used for a VMC calculation on a coronene molecule.
The parallel efficiency on 64 Summit nodes (2688 cores) is above 99.9\%.
This scaling is representative of what one should expect in optimization, DMC, and excited state calculations (i.e., all types of calculations).
Our flexible parallel implementation thus does not seem to have any disadvantage over more standard approaches using MPI. 

\subsection{Graphical processing unit acceleration} \label{sec:GPU}

\begin{figure}
    \includegraphics{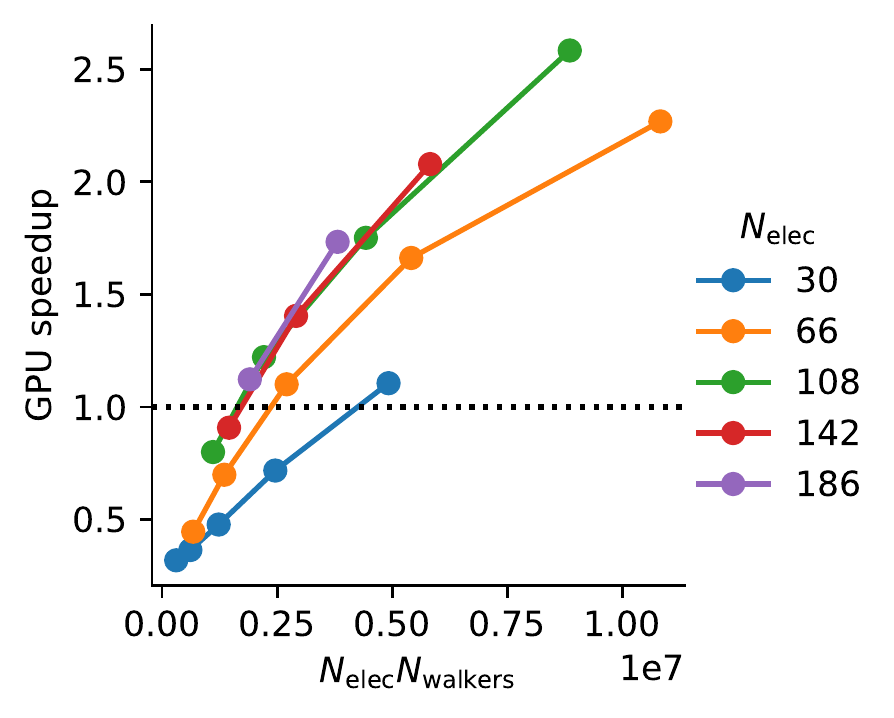}
    \caption{GPU speedup ($t_{\rm CPU} / t_{\rm GPU}$) versus the product of the number of electrons and number of walkers (amount of work). Comparisons are run on a single Summit node, using all 42 available CPUs in both cases, and all 6 GPUs for the GPU case. For large enough problem size, the GPU speedup depends only on the amount of work available.}
\label{fig:gpu_speedup}
\end{figure}

\PyQMC runs on CPUs and GPUs using the same code paths.
GPU cabability is implemented through the \texttt{CuPy} library,\cite{okuta_cupy_2017} which is used as a drop-in replacement for NumPy.
Currently, wave function evaluation and Ewald summation, which are computationally intensive, run on GPU when available and return arrays on CPU.
VMC, DMC, and other algorithmic-level functions are coded entirely on CPU; implementing new algorithms does not require any extra interfacing to make use of GPU resources.

GPUs are massively parallel computing devices that can greatly speed up calculations, but only when given a sufficient amount of computational work to perform.
Fig~\ref{fig:gpu_speedup} shows the GPU speedup (ratio of CPU time to GPU time) versus $N_{\rm walker}N_{\rm elec}$ for a sequence of hydrocarbon molecules: benzene (30 electrons), anthracene (66 electrons), coronene (108 electrons), ovalene (142 electrons), and hexabenzocoronene (186 electrons).
The calculations used correlation-consistent effective core potentials and corresponding VDZ basis sets for both H and C atoms from pseudopotentiallibrary.org.\cite{bennett_new_2017, annaberdiyev_new_2018}
Each calculation was carried out on a single node of the Summit supercomputer at Oak Ridge National Laboratory.
For sufficiently large numbers of electrons (about 60-100), the speedup collapses onto a single line which only depends on $N_{\rm walker}N_{\rm elec}$, approximately the amount of work given to the GPU.

We believe that there could be improvements to the GPU performance of the code by porting more of the code from CPU to GPU.
In particular, \PyQMC uses \PySCF's functions to evaluate the atomic orbitals on the CPU.
For the molecules shown in Fig~\ref{fig:gpu_speedup}, the atomic orbital evaluation takes up 15-20\% of the time, meaning that the GPU speedup in these tests is limited to a maximum of five or six, even if it performed the work in zero time with zero latency.
In the future, we are thus targeting this bottleneck to achieve better GPU speedups.

\section{Conclusion} \label{sec:conclusion}

\PyQMC is a production-level, feature-complete, and state-of-the-art QMC implementation linked with \PySCF.
Because \PyQMC is implemented entirely in Python, it is extremely flexible and modular.
Similarly to \PySCF for standard quantum chemistry methods, \PyQMC is aimed at both production level calculations and development of new methods.  
Just within our group and others, these features have already led to new algorithmic developments.\cite{pathak_light_2020, pathak_excited_2021, yuan_quantification_2022, liInitioCalculationReal2022}
\PyQMC is licensed under the MIT license\cite{MITlicense_OSI,MITlicense_SPDX,MITlicense_origin}, and is thus freely available to download and modify. 
Other groups are free to build on the base implementations laid out here.

Python's high level of abstraction greatly reduces the human time required to customize implementations and develop new ideas.
The library ecosystem is well-developed, including libraries for scientific computing (\texttt{NumPy},\cite{harris2020array} \texttt{SciPy}\cite{2020SciPy-NMeth}), data I/O (\texttt{h5py}\cite{h5py}), parallelization (concurrent, \texttt{MPI for Python}\cite{dalcin_mpi4py_2021}, \texttt{Dask}\cite{dask}), and GPU execution(\texttt{CuPy}\cite{okuta_cupy_2017})
\PyQMC is written in such a way that almost all computationally intensive tasks are actually executed in compiled C or Fortran code provided by one of those libraries, so that the performance is competitive with packages implemented completely in compiled languages while code can be written at high level.

\begin{acknowledgments}
We thank Scott Jensen for helping to proofread the manuscript.
Support from the U.S. National Science Foundation via Award No. 1931258 is acknowledged for development and integration of \PyQMC into \PySCF, and in particular support of W.W. and L.K.W.
Y.C. was supported by the U.S. Department of Energy, Office of Science, Office of Basic Energy Sciences, Computational Materials Sciences Program, under Award No. DE-SC0020177.
Implementing GPU compatibility used resources of the Oak Ridge Leadership Computing Facility at the Oak Ridge National Laboratory, which is supported by the Office of Science of the U.S. Department of Energy under Contract No. DE-AC05-00OR22725.
Additional testing of GPU compatibility used HPC resources of the SDumont supercomputer at the National Laboratory for Scientific Computing (LNCC/MCTI, Brazil).
This work made use of the Illinois Campus Cluster, a computing resource that is operated by the Illinois Campus Cluster Program (ICCP) in conjunction with the National Center for Supercomputing Applications (NCSA) and which is supported by funds from the University of Illinois at Urbana-Champaign.
\end{acknowledgments}

\bibliography{ref.bib}

\end{document}